\pdfoutput=1
\documentclass[preprints,article,accept,moreauthors,pdftex]{mdpi}

\firstpage{1} 
\pubvolume{1}
\issuenum{1}
\articlenumber{0}
\pubyear{2021}
\copyrightyear{2020}
\datereceived{} 
\dateaccepted{} 
\usepackage{sistyle}
\usepackage{comment}
\datepublished{} 
\hreflink{https://doi.org/} 

\usepackage{tikzsymbols}

\Title{Can we detect the quantum nature of weak gravitational fields?}

\TitleCitation{Can we detect the quantum nature of weak gravitational fields?}

\Author{Francesco Coradeschi $^{1}$\orcidA{},
        Antonia Micol Frassino $^{2}$\orcidB{},
        Thiago Guerreiro $^{3}$\orcidC{}*,
        Jennifer Rittenhouse West $^{4}$\orcidD{},
        Enrico Junior Schioppa $^{5}$\orcidE{}}

\AuthorNames{Francesco Coradeschi, Antonia Micol Frassino, Thiago Guerreiro, Jennifer Rittenhouse West, Enrico Junior Schioppa}

\AuthorCitation{Coradeschi, F.; Frassino, A.~M.; Guerreiro, T.; Rittenhouse West, J.; Schioppa, E.~J.}

\address{%
$^{1}$ \quad Istituto del Consiglio Nazionale delle Ricerche, OVI, Italy; fr.coradeschi@gmail.com \\
$^{2}$ \quad Departament de F{\'\i}sica Qu\`antica i Astrof\'{\i}sica, Institut de Ci\`encies del Cosmos,
Universitat de Barcelona, Mart\'{\i} i Franqu\`es 1, E-08028 Barcelona, Spain; antoniam.frassino@icc.ub.edu\\
$^{3}$ \quad Department of Physics, Pontifical Catholic University of Rio de Janeiro, Brazil; barbosa@puc-rio.br \color{black}\\
$^{4}$ \quad Lawrence Berkeley National Laboratory, Berkeley, CA 94720, USA; jennifer@lbl.gov\\
$^{5}$ \quad Dipartimento di Matematica e Fisica ``E. De Giorgi'', Universit\`a del Salento, Lecce, Italy and Istituto Nazionale di Fisica Nucleare (INFN) Sezione di Lecce, Italy; enrico.schioppa@unisalento.it}

\corres{Correspondence: barbosa@puc-rio.br,enrico.schioppa@unisalento.it,antoniam.frassino@icc.ub.edu}

\abstract{A theoretical framework for the quantization of gravity has been an elusive Holy Grail since the birth of quantum theory and general relativity. While generations of scientists have attempted solutions to this deep riddle, an alternative path built upon the idea that experimental evidence could determine whether gravity is quantized has been decades in the making. The possibility of an experimental answer to the question of the quantization of gravity is of renewed interest in the era of gravitational wave detectors. We review and investigate an important subset of phenomenological quantum gravity, detecting quantum signatures of weak gravitational fields in table-top experiments and interferometers.}


\begin{document}

\section{Introduction}
\label{sec:intro}
Constructing a theory that unifies quantum mechanics (QM) and general relativity (GR) has been a nearly century-long effort that continues to this day.  Even with great advances in theoretical quantum gravity, we still do not have a complete solution. Perhaps as a consequence of the enormous difficulty of this endeavor, the critical role that experimental physics could play in the field of quantum gravity was realized early on - a role of increasing interest with the first observations of gravitational waves (GWs) in 2015 \cite{Abbott_2016,Abbott_2016a,Abbott_2016b,Abbott_2017}.

Prior to the 2016 GW discovery papers, proposals for experimental probes of quantum gravity included gamma-ray bursts \cite{Amelino-Camelia:1997ieq}, Michelson interferometers of laboratory scale \cite{Hogan:2010zs}, ultra-high energy cosmic rays and colliders \cite{Hossenfelder:2003jz}, gravitons in hadron collider signatures \cite{Mirabelli:1998rt}, experimental inconsistencies with an alternative to quantum gravity (the semiclassical Einstein equations) \cite{Page:1981aj}, the running of the gravitational coupling $G$ \cite{Hamber:2004ew,Hamber:2009zz}, quantum corrections to gravitational scattering \cite{Donoghue:1994dn,Burgess:2003jk}, molecular interferometry \cite{Carlip:2008zf} and many others, spanning both model dependent space (e.g. string theory or loop quantum gravity dependent) and model independent parameter space. 

From 2016 and onward, an increase in the already widespread interest in detecting signatures of quantum gravity was seen in a growing number of new (or renewed) experimental solutions including those - like interferometers - that could detect possible weak signals in the already weak realm of GW. In fact, although GR correctly explains all current GW observations \cite{Abbott_2017a,Callister_2017,Abbott_2018} and laboratory tests of gravity \cite{Sushkov_2011}, it is still possible that GW will be a window into signatures of the quantization of gravity~\cite{Calcagni_2019}. However, GW observatories like LIGO and Virgo are not the only candidate detectors. Cavity opto-mechanical systems offer complementary windows to astrophysical and spectroscopic measurements \cite{Bonaldi_2020}. Recent proposals in phenomenological quantum gravity range from quantum gravity in electromagnetic cavities \cite{GUERREIRO:2020CQG}, quantum gravity in gravitational wave detectors \cite{Parikh:2020A, Parikh:2020B}, quantum table-top experiments in the lab \cite{Carney_2019}, quantum gravity induced quantum entanglement \cite{Krisnanda:2019glc} to interferometers with rotational sensitivity \cite{Hogan:2016tkp,Richardson:2020snt} and those sensitive to quantum space-time geometry \cite{Holometer:2016ipr,Holometer:2017ovp} (for conjectured holographic quantum geometry effects), to name but a few.  We begin  by defining the approach and scope of the experimental proposals we review in this work.

\section{Quantum Gravity regimes}
The detection of (potential) quantum signatures in gravitational interactions is made especially challenging by the intrinsic characteristics of gravity itself: the extreme weakness of the interaction relative to the other fundamental forces, and -- more importantly -- the dual nature of the gravitational field, which is, in our current understanding, not simply a building block of Nature but defines the structure of space-time itself. This makes even the logical definition of a "quantum gravity problem" nontrivial. 

For these reasons, it is useful to introduce a coarse-grained classification of potential quantum gravity (QG) effects which could lead to experimentally detectable signatures. Following an early leader in this field \cite{Amelino-Camelia:2008aez}, we can identify three different types of QG effects:
\begin{enumerate}
    \item \label{it:RealQGeffects}\textbf{The strong-field QG regime}: this is the domain of quantum black-holes and more generally of strong QG interactions. In these situations, we expect our current theoretical descriptions to be completely inapplicable. Experimental handles on this regime are hard to come by, although it has been done in, e.g. black hole ringdown signatures (``echoes'') of black hole area quantization \cite{Cardoso:2019apo} and discrete mass spectra of primordial black holes \cite{Bekenstein:1995ju}.
    \item \label{it:EFTeffects}\textbf{The perturbative QG regime}: long-distance QG effects are of interest in this regime, staying well below the energies in which gravity becomes as strong as the other interactions. Here we expect QG to introduce small corrections to observable quantities -- corrections which can possibly be detected at sensitive-enough experiments.
    \item \label{it:STeffects}\textbf{The quantum spacetime regime}: for this type of effect, the quantum behaviour of the gravitational field is not directly examined; rather, one takes into account the fact that the structure of spacetime at the typical QG scale (which need not necessarily be the Planck scale) should be drastically affected by the quantization of gravity, typically introducing some effective discreteness or a "fuzziness" in space and time variables, and affecting fundamental symmetries such as Poincaré and CPT invariance. These effects, even if very small, can have observable consequences at scales much below the QG scale.
\end{enumerate}

The main focus of this review is on the perturbative regime, Type \ref{it:EFTeffects}. The reason is twofold: first, there are interesting recent developments in this direction, which suggest new potentially observable effects. The second reason is that these kinds of effects are under the best control from a theoretical point of view, allowing us to make rather solid model independent predictions. Indeed, effects of Type  \ref{it:RealQGeffects} are by necessity strongly model dependent, and unfortunately we currently have many partial candidate models and no complete one. Effects of Type \ref{it:STeffects} are arguably less model dependent, but still require some kind of parametrization of the structure of spacetime at the QG scale. By contrast, perturbative effects can be reliably described in the language of Effective Field Theories (EFT); it has long been postulated \cite{Donoghue:1994dn} that there are no fundamental obstructions in describing gravity as a nonrenormalizable EFT using the standard quantum field theory machinery to quantize the Einstein-Hilbert action, and considering small-field fluctuations around a fixed (not necessarily flat) background metric.  It is interesting to note, in passing, that EFT methods can also be used to study high-energy effects as long as they contain a suitable hierarchy of scales, such as extremely high-energy scattering with small angle of deflection, possibly accompanied by emission of gravitational radiation -- see for instance \cite{Ciafaloni:2015vsa,Ciafaloni:2015xsr} and references therein.

Finally, we would like to point out that assessing the quantum nature of gravity experimentally does not  necessarily require detecting a single graviton, in much the same way that the quantization of the electromagnetic field can be observed in a wide number of experiments that do not detect single photons \cite{Lvovsky2015}. The importance of this statement is reinforced if one follows the reasoning proposed by Dyson, who has shown that the detection of a single graviton might be impossible on first principles~\cite{DYSON:2013jra}.  In this work, we take the point of view that quantum nature of gravity will be established if the quantization of the gravitational field in an EFT sense proves to be necessary in order to describe some experimental result.

With our basic approach and scope being set, we can now ask which kind of effects we can expect in the perturbative regime. As mentioned, these can be foreseen on general principles and do not depend on the details of a hypothetical quantum theory of gravity. As we shall see, a number of publications 
have shown that such effects can be divided into at least three categories: 
\begin{enumerate}
    \item quantum gravitational fluctuations that behave as an additional source of (irreducible) noise on detectors (e.g. interferometers);
    \item effective signals induced by hypothetical non-classical states of gravitational radiation;
    \item non-radiative effects. 
\end{enumerate}
It should be noted that, despite such effects being calculable, the discussion about their measurability (either because of limitations of current/future technologies or due to first principles) has to be assessed separately on a case-by-case basis.

This work contains a narrow selection of topics and references. We focus on explicit effects of the gravitational interaction, including the interaction with a GW detector, as opposed to other phenomenological effects \cite{Pikovski_2012, Bonaldi_2020, Amelino-Camelia:2008aez,Hossenfelder:2009nu}.
 As emphasized in the introduction, much work has been done on these topics over the past years and decades of work have contributed to the field of the phenomenology of quantum gravity as a whole.
Here we try to give an accessible - theoretical and experimental  - overview of quantum signatures of weak gravitational fields in a selection of experiments.

 The review is organized as follows: In Sec.~\ref{sec:Wigner} we give a historical overview of the importance of experimental tests of the quantum nature of gravity. In Sec.~\ref{sec:search_for_signals_of_the_quantization_of_gravity} we argue that the requirement of high energy in order to obtain experimental signatures of the quantization of gravity may be exchanged for extreme precision and control over quantum matter. In Sec.~\ref{sec:Entanglement} we explore the use of decoherence for experimental set-ups, with a focus on the topic of classical gravitationally induced noise and decoherence within recent developments in the field. In Sec.~\ref{sec:quantum-Hamiltonian} we present general methods of constructing the quantum Hamiltonian of a gravitational field interacting with a LIGO-like probe. In Sec.~\ref{sec.quantum-signatures-of-GWs} we focus on quantum signatures of gravitational waves in current and future GW interferometers. Finally in Sec.~\ref{sec:conclusions} we present our conclusions.

\section{Quantum mechanics, measurements and gravity}
\label{sec:Wigner}
We begin with the early discussions of the relation between quantum mechanics, measurements and gravity. 
E.~Wigner was among the first to point out that quantum uncertainty can lead to limitations in measurements of spacetime curvature~\cite{Wigner1979, WIGNER:1957RMP}. While this is not quantum gravity per se, it is closely related to it. As we shall see later, it gives far-reaching hints on how a unified theory of gravity and quantum mechanics should (or shouldn't) look under specific conditions.  Wigner deserves at least partial credit in our view, for having seeded many important developments in the phenomenological quantum gravity direction. 

\begin{figure}
    \centering
    \includegraphics[width =0.7\textwidth]{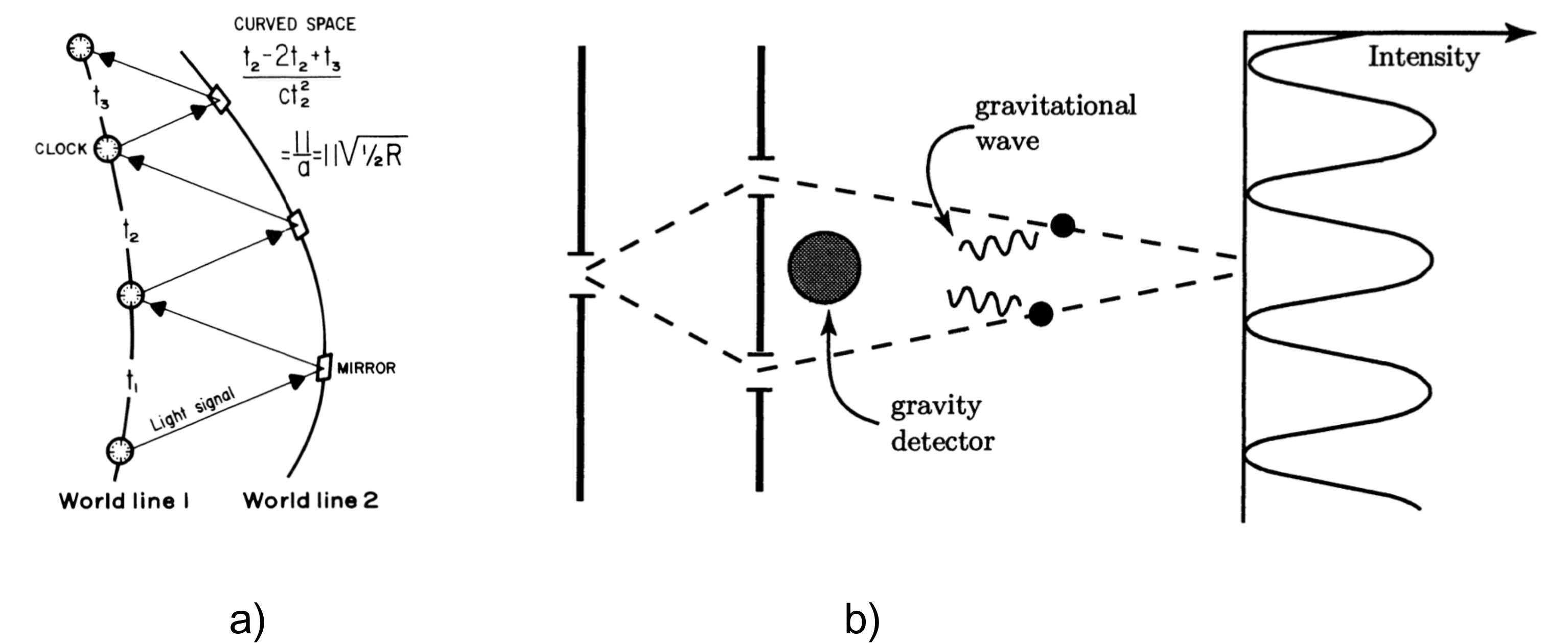}
    \caption[]{a) Wigner (image from~\cite{WIGNER:1957RMP}). b) Feynman (image from~\cite{Feynman2002}).}
    \label{wigner_fig}
\end{figure}

In his 1957 paper, Wigner derived simple relations linking the temporal separation between events measured by clocks to their spatial separation ~\cite{WIGNER:1957RMP}. It is interesting and useful to follow the reasoning behind it. Wigner starts from the argument that the proper measurement of distances in a relativistic theory is not achieved by means of yardsticks or rulers, but by means of clocks and light signals. For example, in a flat metric setup, and after having fixed the reference frame, one considers a first clock, located at event $x_1$, emitting a light signal which takes a time $t$ to reach a second clock at event $x_2$. Immediately after receiving such signal, the second clock emits a new light pulse that will reach the first clock after a time $t'$. Wigner's calculation shows that the space-like distance $D$ between $x_1$ and $x_2$ is given by 
\begin{equation}\label{eq:Wigner1}
    D=c\sqrt{tt'},
\end{equation}
where $c$ is the speed of light.  Although there have been concerns that his 2-dimensional spacetime picture might be ``too narrow'' to conclude that Eq.~\eqref{eq:Wigner1} holds generally \cite{Amelino-Camelia:2007luk}, we use his derivation as it appears in the original paper after cross-checking it; it appears sensible within his specific constrained 2D setup, with the caution that it may not be a general result.  Extending his argument to a non-flat metric, and using the same measurement principle (clocks and light rays), he shows how the components of the Riemann curvature tensor can be directly inferred from measurements of time. For instance, one gets
\begin{eqnarray}
\dfrac{1}{11} \left(\dfrac{t_{1} - 2t_{2} + t_{3}}{t_{2}^{2}}\right) = \sqrt{\dfrac{R_{0101}}{2}},
\label{Wigner}
\end{eqnarray}
when the intervals $ t_{1}, t_{2}, t_{3} $ are defined according to Figure~\ref{wigner_fig}a.
The surprising depth of equation~\eqref{Wigner} is in bringing a component of the Riemann curvature tensor, a powerful mathematical description of curved spacetime, into direct contact with  intuitive and easily understood measurements of the physical world.

From here, Wigner demonstrated that quantum uncertainty can lead to limitations in measurements of the spacetime curvature. For example, a derivation of a lower bound on the mass of a clock $M_c$ which can measure time to a given accuracy is provided. The argument is as follows: since quantum mechanics imposes constraints on how accurately time intervals can be measured, it also limits how accurately one can measure the curvature of spacetime. This is not unrelated to Dyson's considerations on the potential impossibility of directly detecting a single graviton \cite{DYSON:2013jra}. As Wigner concludes, ``\textit{(...) the curvature at a point in spacetime cannot be measured at all; only the average curvature over a finite region of  spacetime can be obtained}''~\cite{WIGNER:1957RMP}. 

Prior to Wigner's work, Osborne investigated the measurement of curvature in one solution of Einstein's equations, the Schwarzchild geometry, using quantum mechanical test particles \cite{Osborne1949}. The conclusion was that quantum mechanics imposes limitations to the accuracy of spacetime measurements and that  ``\textit{in any theory that attempts to unite quantum theory with the general theory of relativity, the relation between the metric and the energy-momentum tensor (Einstein's equations) must appear only in the large and in a statistical sense.}''~\cite{Osborne1949}.

These early considerations by Wigner, Osborne and Dyson highlight the difficulties of constructing a unified theory of quantum mechanics and general relativity. They also reinforce a most important point: the role of experimental tests on the quantum nature of gravity. 

A primary branch of quantum gravity assumes that gravity will be  quantum mechanical. As discussed by Feynman in the \textit{Feynman Lectures on Gravitation} \cite{Feynman2002}, one can make the argument that the world cannot be ``\textit{one-half quantum}'' and ``\textit{one-half classical}.'' Before constructing a theory of quantum gravity, the problem might reside in the proper definition of a starting point. Feynman himself takes the practical (``shut-up and calculate'') approach that the characteristic feature of \textit{quantum} is the need for \textit{amplitudes} describing different processes. 
A particle going through a double slit experiment, illustrated in Figure \ref{wigner_fig}b, has amplitudes associated to the paths traversing the left and right slits, which must accompany the  amplitudes for the gravitational field associated to the particle being sourced at two different spatial locations. 
This was perhaps the first time that gravitationally-induced decoherence was hinted at, a subject we will address later (Sec.~\ref{sec:Entanglement}).

When dealing with experiments attempting to reveal the quantum nature of the gravitational field, a practical approach similar to Feynman's for defining what is meant by \textit{quantum gravity} is desired. We reiterate that investigating the quantum nature of gravity, in this work, implies quantizing the gravitational field in the sense of an effective field theory.  If describing a certain experiment requires the use of creation and annihilation operators of gravitational modes or (small) fluctuations of the gravitational field at zero temperature, then we consider that to be a quantum gravity experiment, i.e. an experiment that can probe the quantized nature of gravity.  In addition, we will see that whatever the full phenomenology of  quantum gravity is finally going to be, it is at least expected that such a scheme of quantum mechanics interacting with weak gravity, would produce fundamental noise on (GW) detectors. 

We have experimental evidence that this approach may be oriented in the right direction. In fact, such a scheme proved its own reliability in 1975, when Colella, Overhauser and Werner~\cite{COLELLA:1975PRL} obtained experimental results showing that \emph{a gravitational potential coherently changes the phase of a neutron wave function}. This could be regarded as the very first experimental evidence of the interaction of gravity with a quantum mechanical system. Since then, more experiments on gravity-induced effects on neutron wave functions (including ``quantum bouncing ball'' setups~\cite{ABELE2009593c}) have been carried out, perfected or proposed (for an overview, see~\cite{Abele_2012} and references therein). As Abele and Leeb summarize in their review~\cite{Abele_2012}, neutrons are a potentially powerful system to investigate possible departures of gravity from Newton's law. Due to their very small electrical polarizability with respect to other systems (notably atoms), neutrons are sufficiently insensitive to short-range electrostatic forces to allow for probing gravitational forces down to sub-micrometer distances.

\section{The search for signals of the quantization of gravity}
\label{sec:search_for_signals_of_the_quantization_of_gravity}
Often the issue with experiments in which gravity and quantum mechanics are both required to explain the data is the way in which relevant quantities such as the Newton constant $G$, the Planck constant $\hbar, $ and the speed of light $ c $ combine to yield very small numbers. This can be exemplified in the work of Donoghue \cite{Donoghue:1994dn}, in which the leading (one loop) quantum corrections to the Newtonian gravitational potential have been calculated.  Such corrections that have been argued not to be correct \cite{Hamber:2017pli}, but in our opinion follow the perturbative nonrenormalizable models of chiral effective theory \cite{Gasser:1983yg} and as such, may yet match with experiment. The corrected potential reads
\begin{eqnarray}
V(r) = -\dfrac{GM_{1}M_{2}}{r} \left[ 1  - \dfrac{G(M_{1} + M_{2})}{rc^{2}} - \dfrac{127}{30\pi^{2}} \dfrac{G\hbar}{c^{3}r^{2}}   \right].
\label{Donogue_correction}
\end{eqnarray}
The  first correction can be obtained in the post-Newtonian formalism, while the second term, proportional to $ \hbar $ corresponds to the quantum correction. Note that it scales as $ (\ell_{\rm pl} / r)^{2} $ where $ \ell_{\rm pl} = \sqrt{G\hbar/c^{3}} $ is the Planck length. This means the quantum correction only becomes of $ \mathcal{O}(1) $ when $ r \approx \ell_{\rm pl} $, which is far from any conceivable experiment today.

Independently of their detailed form, the quantum mechanical corrections in \eqref{Donogue_correction} are in agreement with dimensional arguments in quantum field theory, 
suggesting that the observation of quantum corrections to gravity requires extremely high (Planck scale) energies. A way out of this problem is to theorize new phenomena which shift the scale of quantum gravity down to experimentally accessible energies. In such cases, Dyson's argument can be overcome and gravitons could be probed at current or future collider experiments. The ATLAS~\cite{ATLAS}, CMS~\cite{CMS} and LHCb~\cite{LHCb} collaborations have been attempting to find signals of gravitons ever since the first proton-proton collisions took place at the Large Hadron Collider (LHC) in 2009. Typically, the LHC experiments would look for new heavy resonances in the distribution of specific kinematic parameters of a selected final state~\cite{RAPPOCCIO2019100027}. Such resonances could then be interpreted as graviton states, by fitting the data with exotic models of new physics, such as scenarios with Kaluza-Klein excitations~\cite{Kaluza,Klein} in warped extra dimensions~\cite{RS1,RS2}. Alas, so far with no success, suggesting that the quantum gravity scale is not as accessible as we would hope, and higher energies will likely be needed to continue these searches.

As an alternative, however, the requirement of high energy could be exchanged for extreme precision and control over quantum matter. It was in this spirit that Marletto and Vedral (MV) proposed a test for quantum effects in gravity using systems that locally interact with each other only via the gravitational field \cite{Marletto2017}.
The MV proposal rests upon a general quantum information argument which we now briefly describe. 
Let $ A $ and $ B $ be quantum spins while $ C $ is taken to be a classical bit. That $ C $ is classical effectively means it has only one possible observable $ \hat{q}^{(C)} = 1_{AB} \otimes Z_{C} $, where $ 1_{AB} $ is the identity operator for the joint system $ AB $ and $ Z_{C} $ is the Pauli matrix for system $ C $ along the $ \hat{z} $ direction (similarly, $X_A$ would be the Pauli matrix along $\hat{x}$ for system $A$, $Y_B$ is the Pauli matrix along $\hat{y}$ for system B, and so on). This means the only observation one can make is to measure the value of the boolean variable describing $ C $, for which we assume possible values of $ - 1 $ and $ + 1 $. Spins $ A$ and $ B $ have observables
\begin{align}
    \hat{q}^{(A)} = (X_{A} \otimes 1_{BC}, Y_{A} \otimes 1_{BC}, Z_{A} \otimes 1_{BC}) \\
    \nonumber \\
    \hat{q}^{(B)} = (X_{B} \otimes 1_{AC}, Y_{B} \otimes 1_{AC}, Z_{B} \otimes 1_{AC}) 
\end{align}
Now suppose systems $ A, B $ and $ C $ start in a separable state. System $ A $ interacts locally with $ C $, and $ C $ interacts locally with $ B $, and assume $ A $ and $ B $  \textit{never directly interact with each other}.
The most general state one can write for systems $ A $ and $ C $, for example, is~\cite{Marletto2017}
\begin{align}
    \rho^{AC} = \dfrac{1}{4} \left( 1_{AC} + \Vec{r}\cdot \hat{q}^{(A)}   + s \hat{q}^{(C)} + (\Vec{t} \cdot \hat{q}^{(A)}) \otimes \hat{q}^{(C)} \right)
\end{align}
where $ \Vec{r}, \Vec{t} $ are real vectors and $ s $ is a real number. A similar state can be written for the joint system $ BC $. Note that both $ \rho^{AC} $ and $ \rho^{BC} $ are separable states\footnote{This can be seen by using the Peres–Horodecki (PPT) criterion on system $ C $ \cite{Horodecki:1996nc,Peres:1996dw}.
}. Now, if $ AC $ and $ BC $ are separable and moreover $ A $ never directly interacts with $ B $ (the locality assumption), the total system $ ABC $ must also be separable as it has been prepared using local operations and classical communication. If the total state is separable, it implies that the joint subsystem $ AB $ must also be separable.
The conclusion is that if $ AB $ are found to be entangled after local interaction with a system $ C $, then this contradicts the assumpation that $ C$ is described by a classical variable, and it must be quantum in the sense that it necessarily has an observable that does not commute with $ \hat{q}^{(C)} $. Note that if  $ C$ is quantum in this sense, then it is possible to entangle $ A $ and $ B $ without ever entangling any of these with $ C $ as shown by Cubitt \textit{et. al.}\cite{Cubbit2003}. 
Therefore, the MV proposal can indicate that there is some quantumness in gravity (if that is chosen to represent system C) but does not allow to obtain its detailed structure.

\section{Entanglement and decoherence}
\label{sec:Entanglement}
The MV argument can be generalized and used to devise another method for detecting the quantum mechanical nature of gravity: let $ A $ and $ B $ be quantum mechanical matter systems while $ C $ describes the gravitational field; if $ AB $ only interact via gravity and one can detect the appearance of entanglement this would imply that the gravitational field must be quantized. 
This idea has been independently used by a number of authors, and proposals for experimental tests of quantum gravity have been put forward using spins \cite{Bose2017} and massive systems \cite{Krisnanda:2019glc, vandeKamp2020, Matsumura2020}. See also \cite{Brandao2020} for related ideas applied to quantum optomechanics experiments.
Atom interferometry also provides interesting techniques for the detection of gravitational waves \cite{Dimopoulous2009}, and gravitational induced entanglement can in principle be detected by employing novel quantum information-based sensing protocols robust to thermal noise  \cite{Carney2021}.
In parallel, several proposals to observe macroscopic superpositions in quantum optomechanics have been made (see for example \cite{Stickler2018, Weiss2021}), and experiments are advancing towards reaching the quantum mechanical regime for massive objects \cite{Whittle2021, Whestpal2021}; it is conceivable that in the near future, experimentalists will be able to build Feynman's original gedanken experiment \cite{Belenchia:2019gcc} as the one shown in Figure \ref{wigner_fig}, and access the gravitational field of quantum matter.

A related experimental context in which quantum mechanics and gravity may intertwine is in the \emph{theory of decoherence}~\cite{Zeh:1970zz,Zurek:1991vd}; see \cite{Bassi2017,Zurek:2003zz} for a general review. Utilizing well established mechanisms for gravitational decoherence expected from the quantization of linearized gravity, Blencowe used effective field theory to estimate the decoherence rate of states of a matter scalar field in superpositions of different energy configurations \cite{Blencowe2013} and later Vedral presented a simple calculation leading to the quantum gravitationally-induced decoherence of a spatial superposition of massive objects in the linear coupling regime \cite{Vedral2020}. 

For a superposition of energy states with difference $ \Delta E $ in the presence of a thermal background of gravitational perturbations at temperature $ T $ the estimated decoherence rate is \cite{Blencowe2013}: 
\begin{eqnarray}
\Gamma \approx \dfrac{k_{B}T}{\hbar} \left(   \dfrac{\Delta E}{E_{\rm pl}} \right)^{2}.
\end{eqnarray}
A single atom, say hydrogen, in a superposition of the $ \vert 1s \rangle $ and $ \vert 2s \rangle  $ states, with $ \Delta E \approx \SI{10.2}{eV} $, in the presence of a thermal background of GWs at $ \SI{1}{K} $ decoheres at the well-known rate of $ \sim \SI{10^{-45}}{Hz} $. 
A state of more than two entangled particles, a Greenberger-Horne-Zeilinger (GHZ) state \cite{greenberger2007going}, composed of $ \sim \SI{1}{g} $ (Avogadro's number) of atoms in a superposition of all being in the $ \vert 1s \rangle$ and $\vert 2s \rangle$ states, under the same GW background, decoheres at a more significant rate of $ \sim \SI{100}{Hz} $. A  $ \SI{1}{kg} $ mass of atoms under the same conditions would decohere at a rate of $ \sim \SI{10^{8}}{Hz} $.
The effect of gravitational decoherence in non-classicality tests of gravity has also been analysed in \cite{Rijavec2021}, while the loss of coherence in matter-wave interference experiments due to contact with a thermal bath of gravitons was studied in \cite{Toros2020}. With the advancements in quantum technologies \cite{Poliono2020} and their applications to quantum-limited experiments \cite{McCuller2021} and fundamental physics tests \cite{Backes2021}, the matter of gravitational-induced decoherence is definitely one important research line worth investigating.

Gravitational decoherence of macroscopic entangled states can also be used as a mean of indirectly probing quantum effects of linearized gravity. Kanno \textit{et. al.} \cite{Kanno2021} suggest observing the \emph{decay of entanglement} between massive mirrors in a GW detector.
To generate entanglement, they propose sending a single photon through the interferometer, creating a delocalized excitation \cite{Monteiro2015}. This path-entangled single photon transfers entanglement to the joint motion state of the interferometer's suspended mirrors, creating a cat-state of the form 
\begin{eqnarray}
\vert \psi \rangle = \dfrac{\vert \xi_{1} \rangle \vert 0 \rangle + \vert 0 \rangle \vert \xi_{2} \rangle}{\sqrt{2}}
\label{entangled_mirrors}
\end{eqnarray}
where $ \vert \xi_{1} \rangle, \vert \xi_{2} \rangle $ are mechanical coherent states and $ \vert 0 \rangle $ is the harmonic oscillator ground state.
Decay of this entangled state due to interaction with a noisy bath of gravitons would then indirectly evidence the quantum nature of the gravitational field.

The state~\eqref{entangled_mirrors} decoheres under the influence of a GW background, which can be measured by the decay of the logarithmic negativity, a measure of continuous-variable entanglement. For example, for a background of squeezed waves 
with a cut-off frequency of $ \SI{10^{9}}{Hz} $, a \SI{40}{km} long interferometer with $  \SI{40}{kg}$ mirrors and harmonic oscillation frequency of \SI{1}{kHz}, Kanno \textit{et al.} estimate a decay rate of $ \sim \SI{20}{s} $. Squeezed states are well known non-classical states arising in the theory of quantum optics, that have been conjectured also for gravitational waves. Their peculiarity is that the variance of conjugate QM operators can fall below the minimum uncertainty limit predicted for each operator, while their product satisfies Heisenberg's inequality.

Kanno \textit{et. al.}'s approach can be related to the stochastic modification of the geodesic deviation equation due to quantum fluctuations of GWs by  Parikh, Wilczek, Zahariade (PWZ) \cite{Parikh:2020A, Parikh:2020B}, albeit here in a quantum mechanical context. If Parikh \textit{et. al} derive a classical Brownian motion due to the noise of gravitons, one could interpret Kanno \textit{et. al.} as the quantum mechanical version of that, in which gravitons induce modifications to quantum dynamics, from a unitary evolution
to an open-system master equation. Such dynamics can be effectively interpreted as an average over realizations of a stochastic process in Hilbert space, through the so-called unravelling of the master equation \cite{Molmer1996}. This can be thought of as the quantum analogue of the relationship between the Fokker-Planck and Langevin equations. 

PWZ's proposal presented in~\cite{Parikh:2020A, Parikh:2020B} is based on the intuition that falling bodies are subject to random - quantum induced - fluctuations, or noise. This noise arises due to the interaction of the falling bodies with gravitons.
In particular, they consider as model of a GW detector two free falling masses whose geodesic separation (along the $x$-axes for example) is being monitored.
As the GW passes by free-falling inertial objects that are moving on geodesics of the space-time, the relative separation of those geodesics will change. 
Classical GR accounts for this changing in time as consequence of the curvature of space-time and the description is given by the geodesic deviation equation:
\begin{equation}
\label{eq:GeodEq}
    \ddot{\xi^{\mu}} = - {R^{\mu}}_{0 \nu 0} \xi^{\nu},
\end{equation}
where $\xi^{\mu}$ is the space-like vector connecting two time-like geodesics. 
In the presence of a gravitational wave, described by the weak gravitational field $g_{\mu \nu} = \eta_{\mu \nu} + h_{\mu \nu}$, the relevant components of the Riemann tensor are 
\begin{equation}
\label{eq:Riemh}
    R_{ \mu 0 \nu 0} = -\frac{1}{2} \ddot{h}_{\mu \nu}\,,
\end{equation}
therefore, inserting Eq.~\eqref{eq:Riemh} in Eq.~\eqref{eq:GeodEq}, one gets the equation that describes how the separation of free-falling particles (or nearby detectors) changes when subject to a GW, that is given by:
\begin{equation}
    \ddot{\xi}(t) = \frac{1}{2} \ddot{h}(t) \xi(t),
\end{equation}
where here we dropped the subscript on $\xi_x$ because we suppose that the $x$-axes describes the arm, and $h$ stands for $h_{xx}$.
The question then is if a generalization of this equation exists when the spacetime metric is treated as a quantum field. 
Supposing that the detector is made of two bodies $M$ and $m_0$ on a background spacetime (with $M \gg m_{0}$ and in the free falling frame) and that the linearized gravitational field is quantized, PWZ give a ``quantum geodesic deviation equation''~\cite{Parikh:2020B,Haba_2021} that is a Langevin-like equation
\begin{equation}
    \ddot{\xi}(t) - \frac{1}{2} \left[ \ddot{h}(t) +  \ddot{\mathcal{N}}_{0} (t) - \frac{m_{0} G}{c^5} \frac{d^5}{dt^5} \xi^2(t)\right] \xi(t) = 0\,,
    \label{PWZ_geodesic}
\end{equation}
where $\mathcal{N}_{0} (t)$ is the noise function.
The induced noise has statistical properties that depend on the gravitational quantum state and it appears to be correlated between nearby detectors.
The quantum state depends on the sources and while for a wide range of gravitational sources the deviations from classical behavior are expected to be small, some exceptions can be found. This will be clarified in section~\ref{sec.quantum-signatures-of-GWs}.

As a final remark concerning classical gravitational induced noise and decoherence, we discussed previously the quantum mechanically imposed limitations to measurements of spacetime; conversely, gravitational decoherence forces us to contemplate that spacetime may also impose limitations to the observation of quantum mechanical phenomena. 
The gravitational field cannot be shielded, something appreciated in the early works of Bronstein \cite{Bronstein2012}. As a consequence, it is impossible to prevent decoherence from happening in a thermal background of GWs. To observe coherence 
of a quantum state, one needs to evolve it in time and measure interference fringes; for a superposition of energy eigenstates given by
\begin{eqnarray}
\vert \psi \rangle = \dfrac{\vert E \rangle + \vert E + \Delta E \rangle}{\sqrt{2}}
\label{energy_superposition}
\end{eqnarray}
it takes at least a time $ \tau \approx \hbar / \Delta E $ to observe one interference fringe \cite{Anandan1990}. If the decoherence time $ \Gamma^{-1} $ is shorter than $ \tau $, the coherence in \eqref{energy_superposition} cannot be observed, since the state decoheres faster than the time of one oscillation. This happens whenever the energy difference $ \Delta E $ satisfies 
\begin{eqnarray}
\Delta E \gtrsim \dfrac{E_{\rm pl}^{2}}{k_{B}T}
\label{decoherence_limit}
\end{eqnarray}
Admittedly, this is a large energy gap for typical quantum mechanical experiments under the influence of a thermal GW background expected to be at a temperature around \SI{1}{K} \cite{Allen1996}. For a system composed of $ N $ two-level subsystems with an energy gap of $ \hbar \omega \approx \SI{1}{eV} $ in a macroscopic GHZ state, gravitational decoherence proceeds faster than coherent oscillations when approaching the extremely large number of $ N \approx 10^{60} $ subsystems, which is good news for ongoing quantum computing efforts.
In the early universe, however, when temperatures were much higher and on the order of the Planck energy, formula~\eqref{decoherence_limit} may have played a role in the dynamics of quantum mechanical states. Gravity in the early universe may have provided an important decohering environment, which leads to the intriguing possibility that gravity can also set limits on the observation of quantum mechanical effects. It would be interesting to investigate whether those limits play any role in cosmology.

\section{The quantum gravitational Hamiltonian of LIGO}
\label{sec:quantum-Hamiltonian}
The observation that it is possible to look for a model of quantum mechanics interacting with weak gravity has been put forward by a number of authors who have analyzed what happens to a full system like the LIGO interferometer if we consider it as a detector for the quantum features of gravity. It is indeed conceivable (or at least worth to test) that these effects might reveal themselves in the form of quantized gravitational waves. The point then is to imagine (or better, following Feynman's suggestion: calculate) what the possible effects would be that can produce an unmistakable quantum signature. 
In the following, we review the setting for these investigations, a Hamiltonian formulation where both the gravitational modes and the probe (the detector, or laser interferometer in the case of LIGO) are considered as systems possessing quantized dynamical degrees of freedom.

The quantum Hamiltonian of a gravitational field interacting with a LIGO-like probe can be derived by following the recipe outlined in~\cite{BUONANNO:2003PhysRevD,PANG:2018prD}, which we will briefly summarize. The starting point is a simplification coming from observing, following Buonanno and Chen~\cite{BUONANNO:2003PhysRevD}, that the dynamics of a signal-recycling interferometer can be mapped onto that of a single one-dimensional Fabry-P\`erot cavity. Such mapping is only approximately correct, as it holds upon assuming that a) the radiation pressure forces act on the end test mass mirror and the internal test mass mirror equally, and b) while the light takes its round-trip inside the cavity, the motions of the two mirrors are negligible. As Pang and Chen show~\cite{PANG:2018prD}, such assumptions are valid for LIGO-like experimental setups. The same authors also show that by properly choosing the reference frame in the transverse traceless (TT) gauge, the probe can be mapped onto a very basic optomechanical system where all the dynamics affects the end mirror only, while the input mirror (which is considered to be infinitely massive) has fixed coordinates. In this convenient coordinate system the input mirror sits at the origin. 
If we call $x$ the axis along the one-dimensional cavity of length $L_0$, then $x=0$ is the position of the input mirror.

At this point, the stage is set to derive a full canonical formulation of a linearized quantum theory of gravity that includes interactions with a quantized optomechanical probe~\cite{PANG:2018prD}. On first approximation, one considers interactions involving only incoming and outgoing gravitons. This means that the formalism is appropriate for studying leading order interactions between the physical polaritazions of gravitons and the probe, but it does not include interactions mediated by virtual gravitons, e.g. the self-gravity of the detector. Within this setup, the full system Hamiltonian is written as
\begin{equation}
    \label{eq:full-H-opt}
    H = H_q^{(0)} + H_{EM}^{(0)} + H_{GW}^{(0)} + H_{ext} + H_{OM} + H_{GW}^{(int)}  ,
\end{equation}
where the first three terms are the free Hamiltonians for the test mass (the moving mirror), the electromagnetic field inside the cavity and the gravitational wave, respectively. The latter is the Hamiltonian derived from the linearized Einstein-Hilbert action. The $H_{ext}$ term describes the external pump: ingoing photons to provide for a constant field amplitude inside the cavity, and ougoing photons for the measurement. Pang and Chen write this term as 
\begin{equation}
    H_{ext} = i\sqrt{2\overline{\gamma}}\left[\hat{a}^\dagger\hat{c}_{x=0}-\hat{a}\hat{c}^\dagger_{x=0}\right] - i \int_{-\infty}^{+\infty}dx \hat{c}^\dagger \partial_x \hat{c}\,.
\end{equation}
where $\overline{\gamma}$ is the effective damping rate of the cavity. In this expression and the following ones, the annihilation and creation operators $\hat{a}$, $\hat{a}^\dagger$ are those of the electromagnetic field, while $\hat{c}$, $\hat{c}^\dagger$ are those of the center-of-mass motion degree of freedom of the mirror.

If one calls $\omega_0$ the natural frequency of the cavity, and considers the possibility that the pump field oscillates slightly off resonance with frequency $\omega_L=\omega_0+\Delta$, an additional term must be added to the Hamiltonian, of the form
\begin{equation}
    H_\Delta = -\frac{\Delta}{2} \left(\hat{\alpha}_1^2 + \hat{\alpha}_2^2\right)\,,
\end{equation}
where
\begin{equation}
    \hat{\alpha}_1 = \sqrt{\frac{\hbar}{2}}\left(\hat{a}+\hat{a}^\dagger\right)\,, \qquad \hat{\alpha}_2 = -i\sqrt{\frac{\hbar}{2}}\left(\hat{a}-\hat{a}^\dagger\right)\,.
\end{equation}
are the amplitude and phase quadratures of the electromagnetic field.

The last two terms of the full Hamiltonian~\eqref{eq:full-H-opt} incorporate the optomechanical and gravitational interactions. Following~\cite{GUERREIRO:2020CQG} and references thereof, for a cavity of length $L_0$ and resonance $\omega_0=n\pi/L_0$, one can write
\begin{equation}
    H_{OM} = -\frac{\omega_0}{L_0}\sqrt{\frac{1}{2M\omega_M}}\hat{a}^\dagger \hat{a} \left(\hat{c} + \hat{c}^\dagger\right)\,,
\end{equation}
where $M$ is the mirror's mass and $\omega_M$ its center-of-mass motion resonance frequency. The derivation of such term descends from a proper coordinate transformation that shifts the optomechanical interaction from the boundary conditions of the electromagnetic field (indeed, the oscillating cavity) to an explicit term in the action, as shown in~\cite{PANG:2018prD}. 

Finally, as far as the gravitational interaction term is concerned, Pang and Chen show how its representation changes according to whether one works in the Newton gauge or in the TT gauge (the physics being the same in both gauges)~\cite{PANG:2018prD}. In the former case, the test mass motion derives from the combination of the radiation pressure and the gravitational strain. In the latter case, the gravitational field interacts directly with the electromagnetic field, and the test mass' motion is only determined by radiation pressure. The interaction Hamiltonian in the TT gauge assumes a simple form if one considers a single polarization state of the gravitational wave (e.g. $\lambda=+$) propagating perpendicularly to the cavity axis~\cite{GUERREIRO:2020CQG}, with gravitational wavevector component along the cavity axis $ k_{x} $ satisfying $  k_{x} L_{0} \ll 1$ \cite{PANG:2018prD}. To obtain it, one starts by expanding the linearized field in the TT gauge in Fourier components as
\begin{equation}
    h_{ij}^{TT} \left(t,\bold{x}\right) = \int \frac{d^3k}{\sqrt{(2\pi)^3}} \epsilon_{ij}^\lambda \left(\bold{k}\right) h_\lambda\left(t,\bold{k}\right)e^{i\bold{k}\cdot\bold{x}}\,,
\end{equation}
where $\epsilon_{ij}^\lambda \left(\bold{k}\right)$ are the tensors for the two polarization states $\lambda=+,\times$, satisfying
\begin{equation}
    \epsilon_{ij}^\lambda\epsilon_{jk}^{\lambda'} = \delta_{ik}\delta_{\lambda\lambda'} \qquad \text{orthonormal},
\end{equation}
\begin{equation}
    \epsilon^\lambda_{ij} k^j  = 0 \qquad \text{transverse},
\end{equation}
\begin{equation}
    \text{Tr}\left[\epsilon^\lambda_{ij}\right] = 0 \qquad \text{traceless}.
\end{equation}
By promoting the Fourier coefficients to annihilation and creation operators satisfying the canonical commutation relations
\begin{eqnarray}\label{eq:commutators_h1}
    \left[\hat{h}_\lambda\left(t,\bold{k}\right),\hat{h}^\dagger_{\lambda'}(t,\bold{k'})\right] &=& \delta_{\lambda\lambda'}\delta^{(3)}\left(\bold{k}-\bold{k'}\right)\,,\\\label{eq:commutators_h2}
     \left[\hat{h}^\dagger_\lambda\left(t,\bold{k}\right),\hat{h}^\dagger_{\lambda'}(t,\bold{k'} )\right]  &=& 0\,,\\\label{eq:commutators_h3}
     \left[\hat{h}_\lambda(t,\bold{k}),\hat{h}_{\lambda'}(t,\bold{k'} )\right] &=& 0\,,
\end{eqnarray}
the interaction term for $\lambda=+$ can be written in operator form as
\begin{equation}
    \hat{H}_{GW}^{(int)} = - \frac{\omega_0}{4} \hat{a}^\dagger \hat{a} \int \frac{d^3k}{\sqrt{(2\pi)^3}} \left( \sqrt{\frac{8\pi G}{k}} \hat{h}_\lambda\left(t,\bold{k}\right) + h.c. \right)\,.
    \label{interaction_hamiltonian}
\end{equation}

Such a setup has been used by Parikh, Wilczek and Zahariade in its Lagrangian form~\cite{Parikh:2020B} and by Guerreiro in its canonical formulation~\cite{GUERREIRO:2020CQG} to derive the effects of gravitons prepared under different quantum states upon the detector. 

Note that PWZ, in deriving the modified geodesic deviation Eq. \eqref{PWZ_geodesic}, concentrate on studying the \textit{noise} properties of quantum GWs. 

The fascinating aspect of studying noise properties 
is that it is reminiscent of Einstein's treatment of Brownian noise aimed at establishing the granular nature of matter, and of the detection of fractionally charged Laughlin quasiparticles through measuring the shot-noise statistics of tunneling currents \cite{Saminadayar1997}. 
In all of these approaches, as in PWZ's, statistical properties of noise reveal quantum features of the object of study. 

The authors proceed to estimate the noise correlators and their influence on the motion of the detector's mirrors for quantized GWs in various states of interest using the technique of Feynman-Vernon influence functionals \cite{Feynman2010}.

Once the action for the gravitational field mode interacting
with a free falling mass $m_0$ whose geodesic separation $\xi$ from a heavier fixed mass $M$ is found, the  corresponding Hamiltonian can be quantized.
The Hamiltonian in this case can be written as 
\begin{equation}
    H_{\xi} = H^{(0)}_{q}+ H^{(int)}[\xi]\,,
\end{equation}
i.e., as a sum of a time-independent free piece and an interaction piece \cite{Parikh:2020B}.
In particular, in order to extract the quantum-corrected equation of motion \eqref{PWZ_geodesic} for the length of the detector arm $\xi$, they study the transition probability between two states of $\xi$ in a time interval $\Delta t$. The Feynman-Vernon influence functional is the tool that allows them to determine the dynamics of a quantum system
interacting with another unobserved quantum system (the path integral version of quantum information's partial trace).

The PWZ analysis considers several quantum states of the gravitational field: vacuum, coherent, thermal and squeezed states. We postpone the discussion of the  phenomenology of such noise sources to the next section.

With the alternative canonical approach \cite{GUERREIRO:2020CQG}, one can transform the Hamiltonian \eqref{interaction_hamiltonian} to a discrete set of modes by introducing a quantization volume $ V $, and explicitly  calculate the time-dependence of the mean and variance of the electric field quadrature defined as $ \mathcal{E} = (a+a^{\dagger})/\sqrt{2} $ for various states of the electromagnetic and GW fields. The time evolution, restricted to a single mode of the GW field, results in an operator of the form 
\begin{eqnarray}
U(t)  = e^{iB(t) (a^{\dagger}a)^{2}} e^{q a^{\dagger} a  (\gamma b^{\dagger} -\gamma^{*} b   )}
\label{Uk}
\end{eqnarray}
with $ \gamma = (1 - e^{-i\Omega t}) $ and $B(t) = 2 q^{2} \left(   \Omega t - \sin \Omega t  \right)   $, $\Omega$ being the frequency of the GW mode,
and where $ b $ ($ b^{\dagger} $) represent the graviton annihilation (creation) operator  as defined in the $\hat{h}_\lambda$ operator appearing in equations~\eqref{eq:commutators_h1}-\eqref{interaction_hamiltonian}.

For the relevant mode of frequency $ \Omega $, the coupling constant is $ q = \omega_{0} f_{1} / \Omega $, and $ f_{1} = \sqrt{8\pi G / \Omega V} $ is the \textit{single-graviton strain}. Note that the first term in the evolution operator \eqref{Uk} corresponds to a \textit{Kerr-like} 
nonlinear interaction for the electromagnetic field, also known as a self-phase modulation interaction in which a beam of light modulates its own phase proportionally to its intensity \cite{Lvovsky2015}, resulting in squeezing and squeezing revivals of the cavity electromagnetic quantum state. 

Revivals of squeezing are a purely quantum mechanical effect \cite{Ma2020}, and their observation would confirm the quantum nature of the gravitational field. The appearance of squeezing and revivals even in the case of a GW field in the vacuum state would signal the effect of vacuum fluctuations of the gravitational field upon matter, much like that predicted by Wheeler in his idea of quantum foam \cite{Wheeler1957}.
Unfortunately, the time scale for a revival is prohibitively large, on the order of $ (\omega_{0}/E_{Pl})^{2} \times\omega_{0}^{-1} $ \cite{GUERREIRO:2020CQG}. This, however, epitomizes the idea that the requirement of extremely high energies to probe phenomena such as the quantum nature of gravity could perhaps be exchanged for extreme precision. 

\begin{figure}[h!]
    \centering
    \includegraphics[width =0.65\textwidth]{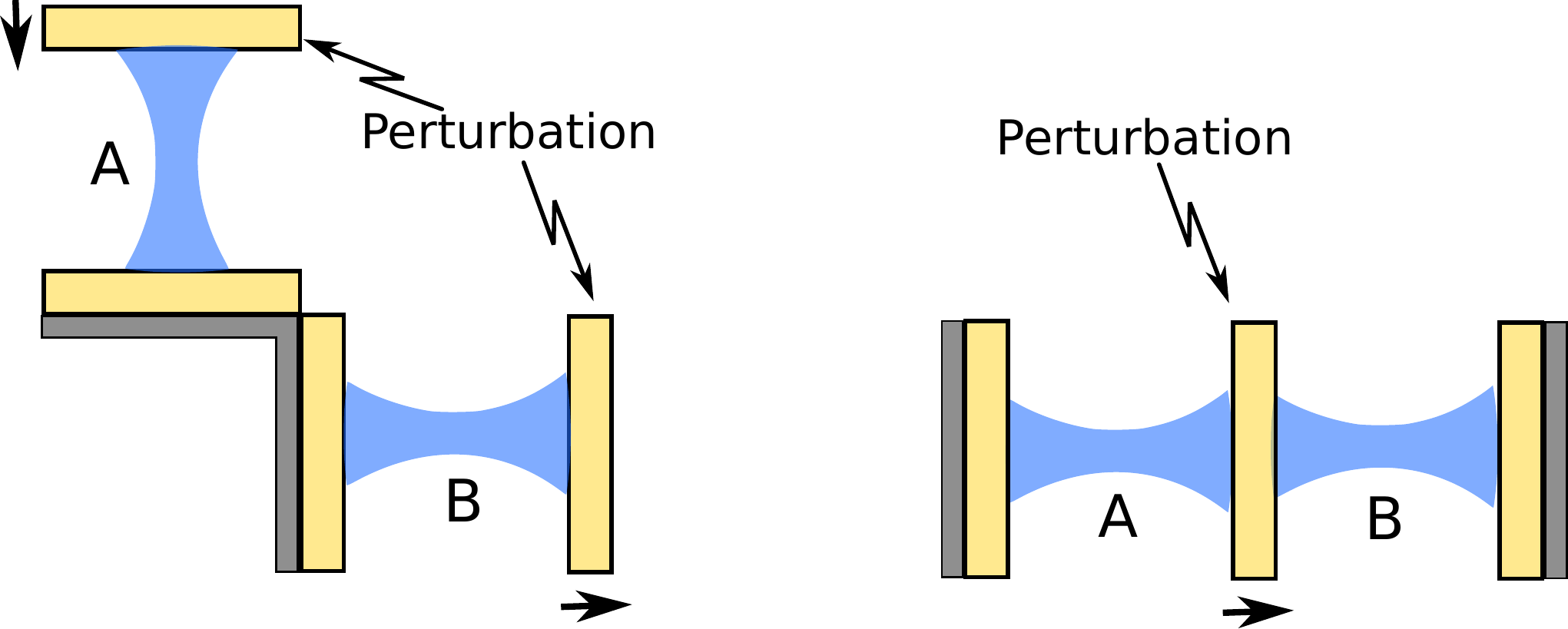}
    \caption[]{Left: two orthogonal cavities. Right: the related situation of a cavity with a mirror in the middle.}
    \label{two_cavities}
\end{figure}

An alternative path to understanding why GW fields in the vacuum state would lead to purely quantum effects detectable in principle using interferometers is outlined in Figure~\ref{two_cavities}.
On the left drawing, two Fabry-P\`erot cavities are set-up with orthogonal axes. A $+$ polarized GW propagating orthogonally to the cavities' axes would then induce quadrupolar motion of the cavities end mirrors according to the indicated arrows. Generalizing \eqref{Uk}, the relevant Kerr-like term would be approximately of the form
\begin{eqnarray}
U_K(t) \approx \exp \left[ \dfrac{\omega_{0}}{E_{pl}} \omega_{0} t (a^{\dagger}_{A}a_{A} - a^{\dagger}_{B}a_{B})^{2}   \right]
\label{two_cavity_evolution}
\end{eqnarray}
where $ a_{A} $ and $ a_{B}$ represent the annihilation operators for cavity modes $ A $ and $ B $, respectively. Being of the second order, the interaction coded in \eqref{two_cavity_evolution} leads to entanglement between the cavity modes $ A $ and $ B$. 
Note that such modes only interact via the dynamical GW degree of freedom, an equivalent situation to the \textit{mirror-in-the-middle} configuration in cavity optomechanics experiments \cite{Brandao2020}, outlined in the right image in Fig.~\ref{two_cavities}. We could hence invoke Marletto and Vedral's argument \cite{Marletto2017}, discussed previously, to claim the quantum nature of the GW field upon the emergence of entanglement between the optical modes. As in the revivals of squeezing, however, such entanglement would take prohibitively long times to build up, which can be traced back to the weakness of gravitational interactions.
In closing this section, we highlight that considering monochromatic GWs, as done here, is a simplification; waves produced in the final stages of binary mergers have frequency chirps, which will lead to corrections to the considerations discussed here. 

\section{Quantum signatures of gravitational waves}
\label{sec.quantum-signatures-of-GWs}

The canonical approach of \cite{GUERREIRO:2020CQG} can also be used to calculate the time-varying cavity electromagnetic field under the influence of a coherent (classical) GW within a quantum mechanical setting. This yields the same result as the classical prediction, that the electric field acquires a time-varying phase, oscillating at the frequency of the GW and with an amplitude proportional to the wave's strain. Alternatively, PWZ also show how, as expected, coherent states recover the phenomenology of classical gravitational waves; in particular: (a)
    there is no way to discern the gravitons that specifically comprise a classical gravitational wave
and (b)
    there is no specific signature of the quantization of gravitational waves emitted by a classical source.

Such conclusions appear to rule out any possibility of ever measuring quantum gravity from gravitational waves emitted "classically," when excluding the presence of quantum gravitational noise. Using different arguments, PWZ also show that thermal states (e.g. cosmic microwave background or the gravitational field sourced by an evaporating black hole) give, at least for today's technology, unmeasurable effects. For example, for an evaporating black hole, any such effects are suppressed by a factor of  $(r_S/r)^2$ where $r_S$ is the Schwarzschild radius of the black hole and $r$ its distance to the detector.

More excitingly, we can also use the same formalism(s) to estimate the effect of GWs in nonclassical states, such as a squeezed vacuum state. Doing that, the result is that the electric field intensity oscillates with an amplitude proportional to the exponential of the squeezing parameter, an enhancement factor found by both Guerreiro~\cite{GUERREIRO:2020CQG} and PWZ~\cite{Parikh:2020B}. Squeezed states are well known from quantum optics, the interest on them relying on the fact that they have no classical analog (such as classical waves in the case of coherent states).  This is not the only place where the importance of squeezing is highlighted in relation to experiments aimed at proving the quantum nature of gravity. For example, Sawyer~\cite{Sawyer2020} has suggested that efficient nonlinear processes capable of converting gravitons into photons are governed by a Hamiltonian capable of producing squeezing.

The central quantity for determining the sensitivity of a GW detector is the power spectral density (PSD) of strain, which is given by the Fourier transform of the auto-correlation of the measured time-dependent strain. PWZ predict the form of the strain noise power PSD 
coming from quantum fluctuations originating from  various quantum states of GWs,  namely the vacuum, thermal and squeezed states \cite{Parikh:2020A}. The predicted PSDs are
\begin{eqnarray}
S(\omega) = \left\{ \begin{array}{ll}
      \dfrac{4 G \hbar \omega}{c^{5}} & \  \mathrm{Vacuum} \\
      \\
      \dfrac{4 G \hbar \omega}{c^{5}} \coth\left( \dfrac{\hbar \omega}{2k_{B}T}  \right) & \   \mathrm{Thermal} \\
      \\
      \dfrac{4 G \hbar \omega}{c^{5}} \sqrt{\cosh 2r} & \  \mathrm{Squeezed}
\end{array} 
\right. 
\end{eqnarray}
where $ T $ is the temperature of a thermal gravitational wave and $ r $ is the squeezing parameter of a squeezed gravitational wave. Note that the dimension of $ S(\omega) $ is \SI{}{Hz^{-1}}, so when comparing with the strain noise of gravitational wave detectors, we are interested in $ \sqrt{S(\omega)} $. 

\begin{figure}[h!]
    \centering
    \includegraphics[width =0.65\textwidth]{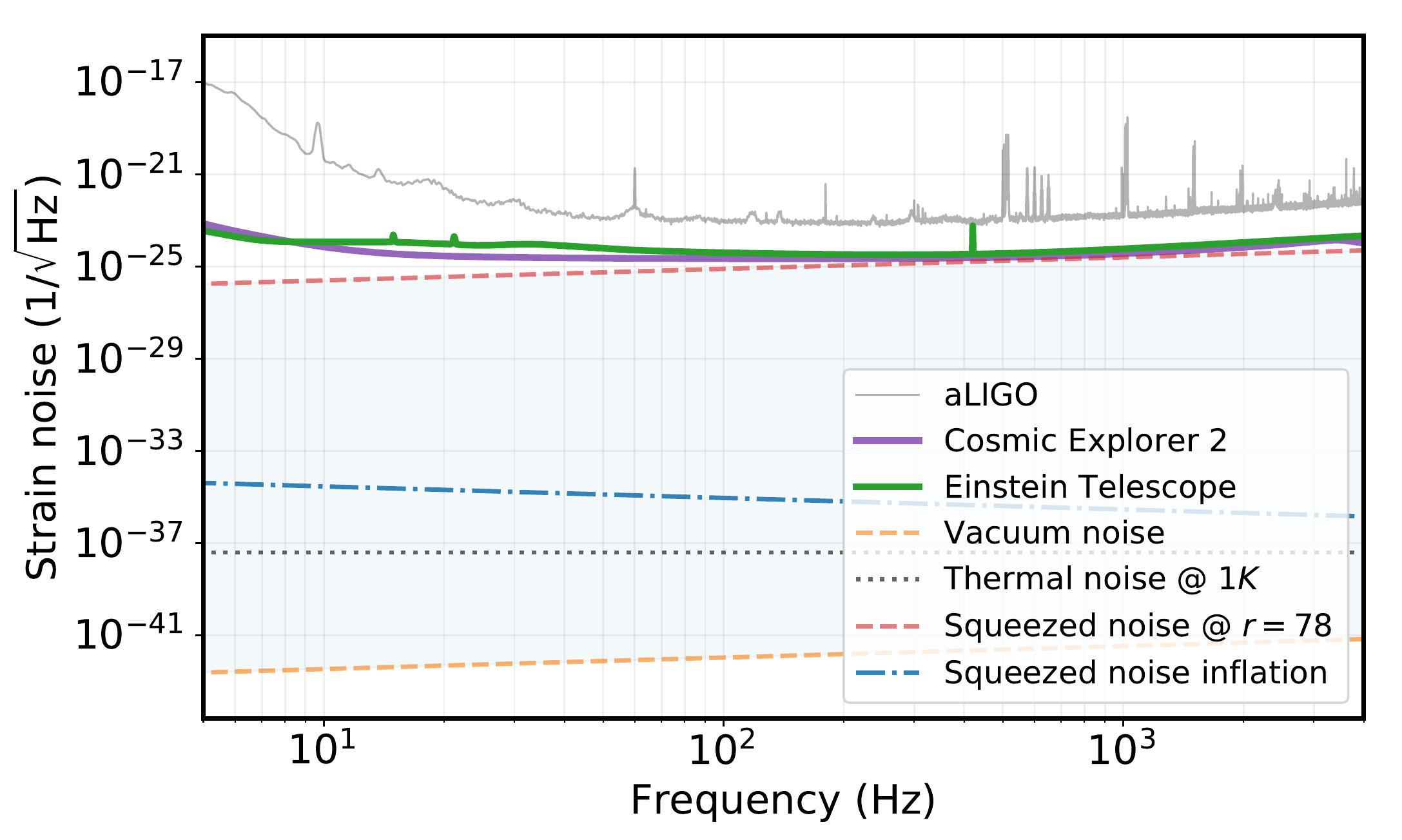}
    \caption[]{Strain plots for current and future GW detectors and the predictions for fundamental strain fluctuations due to the gravitational vacuum, thermal and squeezed gravitational waves.}
    \label{strain_noise}
\end{figure}

In Figure \ref{strain_noise}, the strain noise for the advanced LIGO detector (aLIGO) \cite{DataLIGO}, as well as the future Cosmic Explorer 2 \cite{Evans:2021gyd} and Einstein Telescope \cite{Hild2011} experiments is plotted. 
In the same plot, we show the undetectably small strain noise corresponding to quantum vacuum fluctuations of the gravitational field \cite{Parikh:2020A} (lower dashed orange line), as well as the expected strain noise for a thermal background of gravitons at $ \SI{1}{K} $, the expected current temperature for the cosmic gravitational wave background \cite{Allen1996}. Even though the PSD associated to thermal GW states have an enhancement factor of $ \coth\left( \frac{\hbar \omega}{2k_{B}T}  \right) $, which yields a noise orders-of-magnitude higher than vacuum fluctuations, the effect is still below the expected sensitivity of future experiments. 
Squeezed states, on the other hand, present an exponential enhancement factor in the squeezing parameter, namely $ \sqrt{\cosh 2r} $. This raises the expectation that significant quantum fluctuation effects could be detected in gravitational wave astronomy. The squeezing parameter $ r $ is related to the mean number of quanta $ \langle N \rangle $ in a squeezed state according to~\cite{Scully},
\begin{eqnarray}
\langle N \rangle = \sinh^{2} r\,.
\end{eqnarray}
As an example, a squeezed gravitational wave with $ r \approx 78 $, corresponding to a mean number of gravitons of $ \langle N \rangle \approx  10^{67} $, would be detectable by future GW detectors (see Figure \ref{strain_noise}).

In the terminology of quantum optics, such an amount of gravitons in a squeezed state would correspond to a squeezing of $ \sim 338 $ dB, whereas electromagnetic squeezing experiments produce squeezed states of the electromagnetic field at around $ \sim 10 $ dB \cite{Scully}. In the electromagnetic case, however, squeezing is strongly reduced by loss and decoherence, while in the gravitational case, GWs propagate nearly unperturbed \cite{MTW}. So if there ever was such a violent event capable of producing $ \approx  10^{67} $ gravitons in a squeezed state, future GW experiments could establish the quantum nature of the gravitational field. Moreover, GW astronomy can be used to place limits on sources of strong gravitational squeezing \cite{Chua2015}. 

Squeezed GWs also arise in cosmology; within a simple cosmological model \cite{Allen1996}, the mean number of gravitons emitted in the form of a squeezed gravitational wave of frequency $ \omega $ (as measured today) produced during inflation is 
\begin{eqnarray}
\langle N \rangle = \dfrac{H_{0}^{2} H_{ds}^{2}}{4\omega^{4}} 
\end{eqnarray}
where $ H_{0} $ is the Hubble expansion rate observed today and $ H_{ds} $ can be defined in terms of the scale factor during the de-Sitter phase expressed in terms of comoving time, $ a = \exp(H_{ds} t)  $. 
Kanno \& Soda \cite{Kanno2019} relate this number and consequently the squeezing number to cosmological parameters via the formula
\begin{eqnarray}
\sinh r = \dfrac{1}{2} \left(  \dfrac{ \omega_{1}}{\omega}   \right)^{2}
\end{eqnarray}
where 
\begin{eqnarray}
\omega_{1} = 2\pi \times 10^{9} \sqrt{ \dfrac{H}{10^{-4} M_{Pl}} } \ \SI{}{Hz} 
\end{eqnarray}
Assuming the Hubble parameter $ H = 10^{-4} M_{Pl} $ \cite{Kanno2019}
we obtain the dash-dotted blue curve in Figure \ref{strain_noise}. Interestingly, the PWZ prediction for the noise associated to squeezed states produced by inflation falls somewhat between the unmeasurably small quantum vacuum fluctuations and the expected sensitivity for future GW detectors. This supports the idea that cosmology might be a viable path to uncover quantum gravitational effects \cite{Wilczek2014}.

In any case, the considerations of this work, as well as the results of \cite{Parikh:2020A, Parikh:2020B} and \cite{GUERREIRO:2020CQG} indicate that further studies of nonlinear sources of quantum gravitational waves in strong general relativity should be highly relevant to experimental investigations into the quantization of gravity \cite{Guerreiroplus:2021ig}.

\section{Conclusions}
\label{sec:conclusions}

In this work, we have reviewed recent progress in detecting the quantum nature of weak gravitational fields. Touching on the historical foundations of the field of phenomenological quantum gravity, we have summarized experimental signatures of a possible quantum nature of gravity that can (or cannot!) be measured in the near or far future. We focused on gravity-induced quantum effects on matter (mainly decoherence and entanglement) and on possible signatures at GW interferometers. For the latter, our attention was drawn to the recent work from PWZ, Guerreiro and others, who have shown how we might expect the quantum nature of gravity to affect the noise at GW interferometers. In particular, hypothetical squeezed states of (quantized) GWs are attracting the attention of the community. Future studies may reveal how squeezed GWs might be produced at specific astronomical sources. If the effect is proven to be large enough, the chance of observing it might be closer to our current technological limits than we think.

To conclude, we would like to quote the following extract from Feynman's lectures on gravitation~\cite{Feynman2002} from 1962-63:
\begin{quote}
    Gravitation is so weak that no experiment that we could perform today would be anywhere near sensitive enough to measure gravitational radiation waves, at least, those which are expected to exist from the strongest sources that we might consider, such as rapidly rotating double stars. And the quantum aspect of gravitational waves is a million times further removed from detectability; there is apparently no hope of ever observing a graviton.
\end{quote}
While Feynman was careful when speaking about GWs, leaving room for the experiments of his tomorrow (our today), he appears to close the door on the possibility of detecting the quantum nature of gravity.
However, many of the works reviewed here might give us the key to reopen Feynman's closed door. As put forward by Dyson and Feynman, it seems indeed unlikely that we will ever observe \emph{a graviton}, although indirect detection of a single graviton may provide the way forward \cite{Carney:2021vvt}. However, this does not preclude the possibility that the quantum nature of gravity might eventually be experimentally determined.

\funding{T.G. acknowledges the Coordenac\~ao de Aperfei\c{c}oamento de Pessoal de N\'ivel Superior - Brasil (CAPES) - Finance Code 001, Conselho Nacional de Desenvolvimento Cient\'ifico e Tecnol\'ogico (CNPq) and the FAPERJ Scholarship No. E-26/202.830/2019.  J.R.W. acknowledges partial support from the LDRD program of Lawrence Berkeley National Laboratory, the U.S. Department of Energy, Office of Science, Office of Nuclear Physics, under contract number DE-AC02-05CH11231 within the framework of the TMD Topical Collaboration.  A.M.F. acknowledges  support from the ERC Advanced Grant GravBHs-692951, MEC grant FPA2016-76005-C2-2-P, and AGAUR grant 2017-SGR 754.}

\acknowledgments{
We are thankful to Jorge V. Rocha, Daniel Carney and Herbert Hamber for helpful comments on the manuscript.  We note that in the summer of 2017 the authors participated in the Prospects in Theoretical Physics program (PiTP) at the Institute for Advanced Studies, Princeton, which turned out to be a most inspirational experience. Without that summer school, the authors would have never met, and this work, as well as many hours of exciting fundamental physics discussions would have never happened. Therefore, we acknowledge the "Particle Physics at the LHC and Beyond" PiTP and co-directors Chiara Nappi and Nima Arkani-Hamed  for converging our worldlines.} 

\conflictsofinterest{The authors declare no conflict of interest. The funders had no role in the design of the study; in the collection, analyses, or interpretation of data; in the writing of the manuscript, or in the decision to publish the~results.}


\abbreviations{The following abbreviations are used in this manuscript:\\

\noindent 
\begin{tabular}{@{}ll}
EFT & Effective Field Theory\\
GHZ & Greenberger, Horne and Zeilinger\\
GR & General Relativity\\
GW(s) & Gravitational Wave(s)\\
MV & Marletto and Vedral\\
PSD & Power Spectral Density\\
PWZ & Parikh, Wilczek and Zahariade\\
QG & Quantum Gravity\\
QM & Quantum Mechanics\\
TT & Transverse Traceless\\
\end{tabular}}

\appendixtitles{yes} 
\appendixstart
\appendix


\end{paracol}
\externalbibliography{yes}
\bibliography{references_2}

\end{document}